# Ferroelectric terpolymer films with enhanced cooling efficiency: An integrated approach considering electrocaloric response and dielectric losses

Nouh Zeggai*, [a], Martino LoBue [a], Morgan Almanza [a]

[a] Université Paris-Saclay, ENS Paris-Saclay, CNRS, SATIE, 91190 Gif-sur-Yvette, France

Corresponding author: e-mail: nouh.zeggai@ens-paris-saclay.fr

Declarations of interest: none

**Abstract**

In response to the growing demand for more efficient, and compact refrigeration and energy conversion devices, electrocaloric (EC) P(VDF-TrFE-CFE) (P, poly; VDF, vinylidene fluoride; TrFE, trifluoroethylene; CFE, chlorofluoroethylene) are among the most promising active substances. However, despite their high electrocaloric response, the maximum efficiency achievable over a cooling cycle is hampered by losses. To overcome this major limitation, losses have been reduced by using an electro-thermal poling treatment as well as by controlling the surface roughness. The upper bound of the efficiency computed over a thermodynamic cycle mimicking the working conditions when of an actual cooling device has been increased from 1% to 10% of the Carnot efficiency. This represents a major improvement in enhancing ferroelectric materials for advanced energy applications.

## I. Introduction

Refrigeration, air-conditioning, and heat pump industries are driven by a growing demand for a more efficient, inexpensive, clean, and versatile equipment. According to the International Institute of Refrigeration (IIR), cooling alone represents 15 % of the total world electrical energy consumption, a figure set to grow substantially in the next 20 years due to climate change. Accordingly, research towards more efficient eco-friendly cooling technologies for air conditioning and refrigeration is booming. Solid-state electrocaloric (EC) cooling, which harness reversible temperature change driven by the application of an electric field, represents a promising alternative to vapour-compression systems.

PVDF-based polymers have been studied for applications to electric energy storage [1–3]., electro-strictive actuation [4,5], and EC cooling [6–8]. The outstanding EC response, in terms of adiabatic temperature change, of P(VDF-TrFE-CFE) terpolymer, has been consistently reported for the last few years [7–9]. Since the discovery of the EC effect by Neese in 2008 [10], several cooling devices have been manufactured. However, in spite of the high caloric response of the deployed materials, the reported demonstrator efficiencies are still rather limited, achieving only a few percent of the Carnot efficiency [7–9]. This major shortcoming is mainly to be attributed to the losses taking place within the refrigerant material [11]. Furthermore, losses impose a harsher challenge to efficiency than the limited adiabatic temperature change as the latter can be overcome through suitable device design (i.e. cascading thermodynamic cycles) while the former cannot but be tackled through material improvement (i.e. through actual dielectric loss reduction).

Yet, whereas reports of better EC responses follow one another, losses and their reduction in EC materials received a relatively poor attention.

Material losses are closely connected to the operating conditions, and the significant non-linear behavior they exhibit under application of the high electric fields commonly used in EC cooling applications (e.g., 80 V/μm) [9] [7], makes their understanding, and mastering a major scientific challenge. [9] [9] [9] [9] [9] [9] [8]

As a matter of fact, conduction and the ferroelectric/structural change, namely the main physical mechanisms expected to underlie losses, are significantly influenced by factors like the field strength, the excitation profile (e.g. unipolar or bipolar), and the working frequency [12]. Actually, bipolar profile shows the same EC response than the unipolar one, but exhibits an additional loss due to the entire flipping of the dipoles. Besides, a square voltage profile shows four times more losses compared to a sinusoidal one [11]. For instance, increasing the frequency from 13 Hz to 26 Hz leads to 5 % loss reduction. Therefore, to optimize a device it is crucial to assess losses under the relevant operating conditions getting beyond the standard characterization protocol carried out under sinusoidal excitation at 100 Hz. The typical excitation used in cooling devices involves a unipolar square voltage profile at a frequency of 1 Hz [7] [13].

Here, rather than independently addressing losses, and EC response an integrated approach where the entangled role of the two is studied jointly through a new figure of merit . The approach proposed is similar to the one used for studying energy storage applications, where the energies stored per unit volume and the charge/discharge efficiency ($\eta_{storage}$) are jointly taken into account. In cooling defining a suitable figure of merit for functional materials is a rather uneasy task without taking into account the working conditions specific to each application. A first, and rather natural figure of merit to assess caloric materials, uses the two intrinsic material properties defining the actual intensity of the caloric response. These are the maximum adiabatic temperature change, and the maximum isothermal entropy change. A second, more application-minded step takes into account the work needed to obtain a given caloric response. As in the former case this may focus on obtaining the maximum temperature change, or the maximum isothermal entropy change. The latter has been commonly used within the EC materials community by defining a "cooling efficiency" as the ratio between the heat Q exchanged in isothermal conditions and the work needed to obtain the associated entropy change W, namely $\eta = |Q/W|$. As argued in [14], this figure of merit has the unquestionable advantage to be purely material dependent getting rid of all the details related to the specific working cycle adopted in a given device. However, to address the efficiency of a cooling device deploying a given refrigerant a thermodynamic cycle exchanging heat with two reservoirs at different different temperature must be considered. As its coefficient of performance depends on the temperature of the reservoirs, we refer to the relative efficiency (i.e. the COP of our cycle normalized with the Carnot efficiency). The COP is computed over a thermodynamic cycle made by two irreversible transformations performed in adiabatic conditions (i.e. without heat exchange with the reservoirs), where the measured losses are added, and two isothermal transformations where the entropy production associated with the losses is exchanged in order to recover the initial state (i.e. cyclic transformation). This, as described in detail in [11] ,

represents a Carnot-like refrigerator hampered by the losses directly measured on the refrigerant material. Thence, the reference cycle used here can be considered, in terms of efficiency, as the best cooling cycle achievable using a refrigerant exhibiting the observed losses. This approach has the advantage of addressing a cyclic transformation similar to the ones used in cooling applications, but closer to a Carnot's refrigerator and thence giving an estimation of the best COP achievable when using a given refrigerant notwithstanding the details of the design specific working cycles (e.g. Ericsson, Brayton, etc.).The material ability for cooling, taking into account both the caloric response, and the dielectric losses, are addressed at the same time through the efficiency relative to Carnot ($\eta_{cooling}$ in Eq. 2) computed over a relevant thermodynamic cycle. This amounts extrapolate the maximum COP achievable with a given refrigerant substance from the directly measured material properties. The main goal of this work is to make the case for using the cooling efficiency $\eta_{cooling}$ defined here as a standard figure of merit for EC materials [11] [15].

To show the usefulness of the proposed approach, and to explore, at the same time, new methods towards dielectric loss reduction in EC materials two original roads to get more efficient EC refrigerants are presented and discussed in detail.

The first consists in controlling the surface micro-structure through precise deposition conditions, thereby reducing losses associated with charge injection. The properties of two samples, obtained through deposition in $N_2$ and ambient air, and exhibiting different degrees of roughness are compared.

The second consists in irreversibly modifying the polymer structure (i.e. chain organization, crystalline phase) through an electro-poling treatment consisting in the application of a high periodic unipolar field. Two poling methods have been investigated: one using a moderate unipolar electric field at 30 V/µm assisted by a heating above room temperature (Electrothermal-poling), the other where a progressive increasing of the field up to 70 V/µm is applied while keeping the film at room temperature (Electro-poling). Both methods improved EC response (up to 30% $\Delta T_{adia}$ increasing), and reduced ionic and ferroelectric losses resulting in an efficiency increasing up to 10% of Carnot one. Moreover, films endurance has been improved too.

Finally, this work presents a consistent measurement of adiabatic temperature change and of the efficiency of an optimized P(VDF-TrFE-CFE) performed under relevant working conditions (voltage profile, frequency, film thickness, etc). These two essential material factors of merit for cooling have been established simultaneously for different electric fields and temperatures. The former allow to address the tradeoff between efficiency and adiabatic temperature change; the latter allows investigating the optimal working temperature range. .

The paper is structured as follows: in part II losses and mitigation methods are briefly discussed, in part III two original methods to reduce the losses are described, in part IV the overall improvement of the treated films is discusses through direct measurement of their adiabatic temperature change, and of their losses. Both measurements are used to extrapolate the upper bound of the film cooling efficiency and to detect their optimal working conditions in terms of excitation field and temperature.

## II. Dielectric losses in EC materials

Losses in P(VDF-TrFE-CFE) are related with the complex structure of the material. As a semi-crystalline polymer, P(VDF-TrFE-CFE) is made of an amorphous part and of different crystalline phases, namely the PE (paraelectric), RFE (relaxor ferroelectric), and FE (ferroelectric). Their appearance and their relative fraction depends on the temperature and on the electric field [16] [17]. Each phase shows rather distinct electrical properties and exerts its influence on the overall structure. When an electric field is applied, the relative fraction of these phases changes, leading to a field induced phase transition, i.e. switching from the relaxor ferroelectric phase/paraelectric phase to the ferroelectric phase [18] [19] . These phase changes are the very origin of the EC effect (i.e. of the field induced reversible temperature change) [20]. Nevertheless, electric field excitation and the response of the material entail losses. For instance, delays or energy barriers accompanying the polarization and phase transition mechanism results in entropy production and eventually in dielectric losses. This can be mitigated by changing the molecular structure and the chain packaging with physical or chemical pinning [21], with the addition of double bound [22] or more generally through the synthesis conditions, i.e. the solvent or the annealing conditions [12] [23].

In addition to the ferroelectric losses, the device operating conditions (i.e. the low frequency unipolar excitation at rather high fields) may induce an increasing of the conduction losses [24]. This is also affected by the material quality (i.e. defect density), as by the electrode properties, by the presence of moisture absorption, etc. As a post-treatment method, an electro-thermal annealing on P(VDF-TrFE-CTFE) for energy storage has been demonstrated as a way for reducing conduction losses (or leakage currents) [25]. This method is particularly efficient at relatively low electric field (30 V/µm), and under a unipolar sinusoidal excitation below 0.1 Hz, a frequency where ions conduction associated to impurities is often predominant. Synthesis conditions also lead to different types of chaines entanglement which can be not stable over time [26] and may lead to some displacement driven by the electric field entailing transient losses along the first cycles operated. Among conduction mechanisms, charge injection is also expected to play an important role in the conduction at high field. In this respect, the choice of the best materials for the electrodes is a key challenge. For instance, it has been shown that using gold instead of aluminum as the electrode material [27] as well as incorporating an aluminum nitride (AlN) layer mitigates charge injection and enhances the insulation properties of P(VDF-TrFE-CFE) [27]. Pedroli et al. implemented a P(VDF-HFP) barrier layer at the interface of P(VDF-TrFE-CTFE), resulting in a conduction losses reduction of almost 70% [28]. Charge injection is correlated with the interface field, which is affected by the film surface roughness [29]. Therefore, mastering the surface roughness is a promising route towards the reduction of losses , especially as it does not need an additional layer that would introduce a thermal load detrimental for EC cooling. Works on improving PVDF film transparency, a material close to the terpolymer studied here, show that controlling air humidity during the deposition is an effective method for reducing the film surface roughness with a reported improvement from 2 µm to 10 nm (rms) [30]. Such an improving of the film smoothness is expected to have a substantial impact on P(VDF-TrFE-CFE) losses.

### III.     Enhancing P(VDF-TrFE-CFE) properties

In this work two original method for enhancing materials cooling ability are presented. The first focuses on controlling surface smoothness through film deposition in controlled atmosphere thereby reducing losses associated with charge injection. The second method uses the application of unipolar electric field to induce an irreversible rearrangement of the chain organization. This treatment, called electro-poling, amounts to a sort of material "training" where an optimal chain configuration exhibiting better functional properties is prepared by the periodic application of a field of amplitude greater than the one to be used under working conditions.

To the best of our knowledge, studying losses reduction using unipolar poling has never been reported before. The results presented here show that P(VDF-TrFE-CFE) structure is irreversibly modified by the periodic application of high electric fields.

The term "poling" in ferroelectric refers to the process used to orient all domains in the same direction and to polarize the material. DC voltage or AC voltage are used for ceramic or P(VDF-TrFE) respectively, showing that different mechanism are at play. The technique presented in this work, in spite of its similarity with poling, shows different and deeper effects on the material structure.

## IV. Experimental part

The schematic in Figure 1 illustrates the synthesis conditions, the poling methods and the characterization techniques, used in this work.

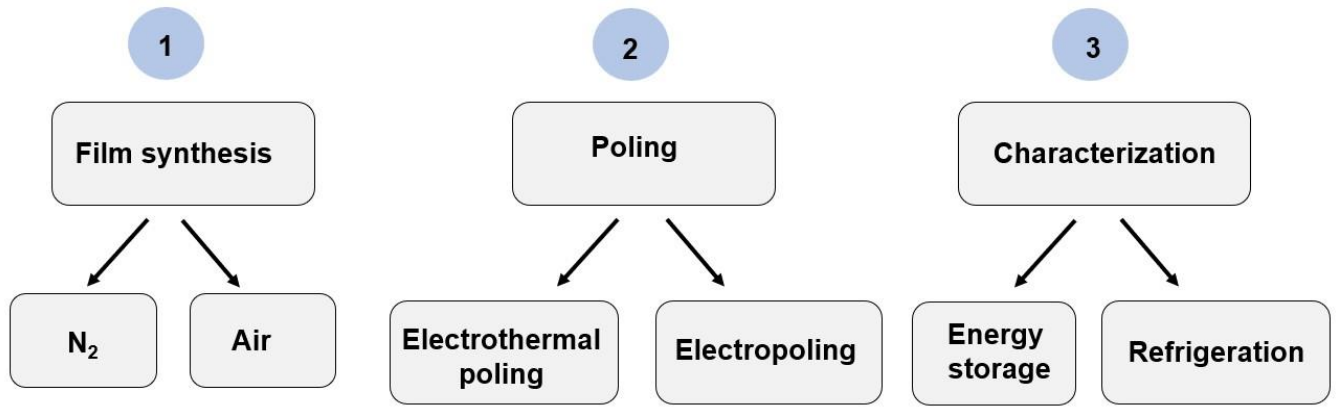

**Figure 1** . Methods for the synthesis, the poling, and the characterization.

### 1. Film fabrication and processing.

Polymer films are prepared using the solution casting method, which involves dissolving P(VDF-TrFE-CFE) terpolymer powder in DMF (Dimethylformamide) solvent at a concentration of 10 *%wt*. This terpolymer contains 66.4% VDF, 24.5% TrFE, and 9.1% CFE, expressed as molar fraction and purchased from Piezotech®. The dissolution temperature was set to 50°C. To modulate the surface state of the film, two deposition processes were used: one under low-humidity environment (deposition in Nitrogen gas) and the other with ambient humidity (deposition in air). A PMMA® box was constructed around the bar coater to keep the gas. The flow of nitrogen is stopped only during the deposition. After casting, the glass plate is left on the plate at 30 °C for 60 minutes to form a film. This process yields a smooth, uniform film. For deposition in air, a similar procedure was followed, with two exceptions, during the deposition, the plate was preheated to 55 °C and the PMMA® box is removed. After deposition, the film was subjected to 60 °C in an vaccum oven for 24 hours to eliminate any remaining solvent. Subsequently, the film was peeled off, and its crystallinity was enhanced through thermal annealing at 102 °C for 2 hours. The obtained film has a thickness of 24 µm ± 1 µm. The roughness of the PVDF films were measured using mechanical profilometer (Dektak). A stylus was then moved across the film's surface (1 µm), recording height variations at regular intervals. To fabricate a capacitor-like structure, rectangular gold electrodes measuring 7 mm x 25 mm (±10 %), and 30 nm in thickness were sputtered on both sides of the terpolymer film (Quorum 300T).

## 2. Electro-poling using electric field and temperature.

Poling the material under specific conditions is aimed at driving the polymer towards a configuration where dielectric losses get minimized keeping, at the same time, a high EC response. The poling process consists in applying an alternating unipolar electric field (0 V to $E_{max}$ and vice versa) at a frequency of 1 Hz. This treatment has been carried out under two different conditions, the ambient electro-poling (E poled) and the temperature electro-poling (E-T poled).

**E poled**: The film is poled at room temperature with a sinusoidal, unipolar electric fields, increasing from 10 V/µm to 100 V/µm with increments of 10 V/µm . Each field amplitude was applied for 20 periods. The sequence showed in (Figure 2.a) is applied 4 times to obtain a stable response. Only the first sequence is stopped at 80 V/µm to avoid premature breakdown.

**E-T poled:** the polymer film undergoes a cyclic unipolar sinusoidal electric field with 30 V/µm for 20 periods, at 30°C then at 35°C and so on up to 50°C. (Figure 2.b).

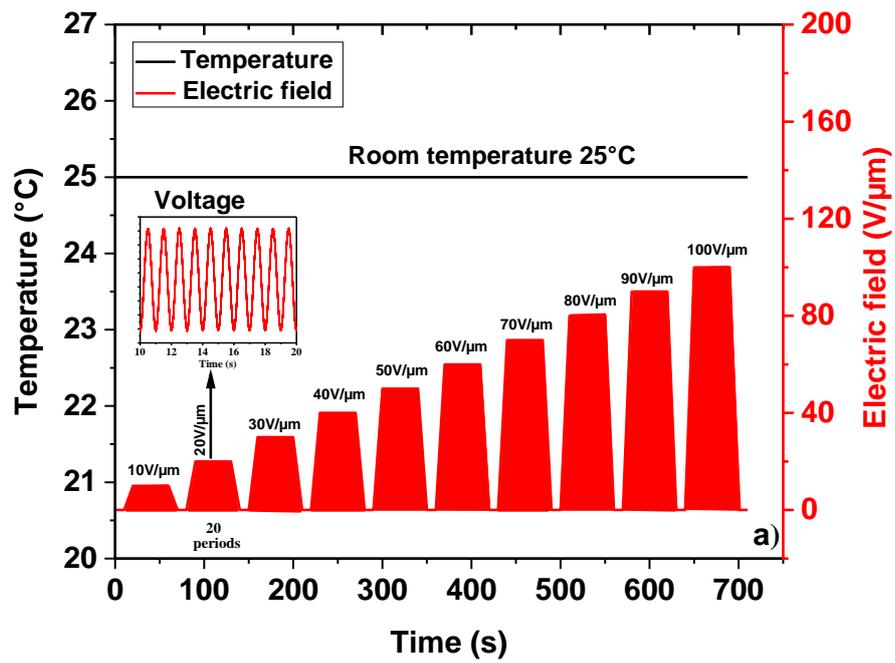

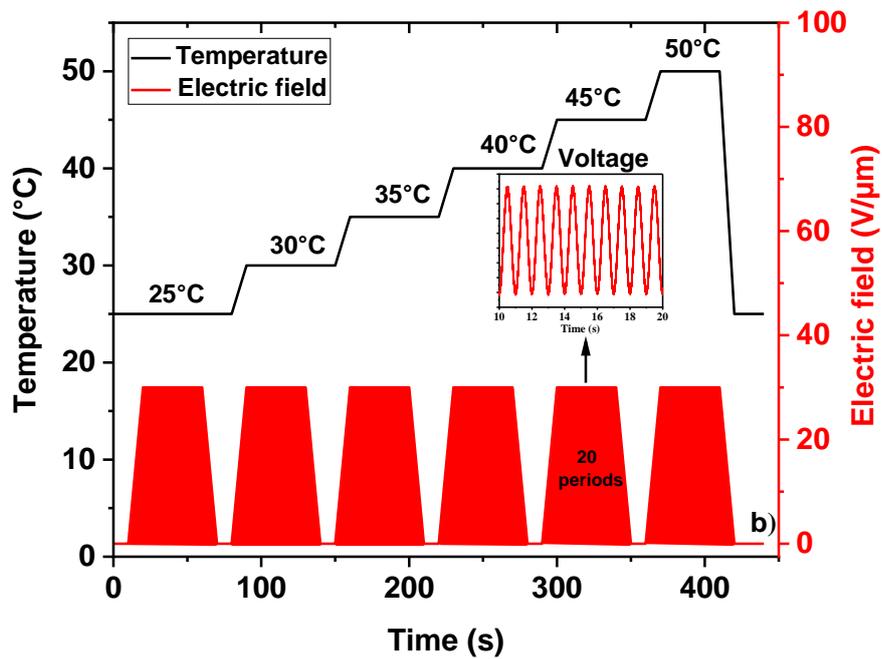

**Figure 2**.Temperature and voltage profiles for, **a)** E poled and **b)** E-T poled treatments.

## 3. Material characterization

The capacitor is powered by a high-voltage amplifier (Matsusada AMP-10B10), with simultaneous measurements of current and voltage to generate polarization-electric field (P-E) hysteresis loops (Figure S 1). The sample is freestanding and the measurement is performed in air. The energy density stored ($E_{storage}$) in the polarization-electric field is integrated from the maximum field to zero (the blue area in Figure S 1). The loss ($W_{loss}$) is represented by the hysteresis loop area (the red area in Figure S 1). To avoid transient effects, loss is calculated over the 20th period using the profile showed in Figure 2.a. From $E_{storage}$ and $W_{loss}$, the charge-discharge efficiency is computed.

$$\eta_{storage} = \frac{E_{storage}}{E_{storage} + W_{loss}} \tag{1}$$

$W_{loss}$ is either measured with a unipolar sinusoidal voltage for studying charge-discharge losses/efficiency or with a unipolar square voltage for evaluating the cooling losses/efficiency. The EC effect was measured directly using a flexible thermistor setup that has been presented in detail elsewhere [11]. The thermistor is kept in direct contact with the terpolymer film. It consists of an 8 μm thick polypropylene (PP) film, coated with an 8 nm thick aluminum serpentine. The resistance of the flexible thermistor, approximately 1 kΩ, has been measured every 50 ms using a 4-point probe setup with a digital multimeter. The thermistor exhibits a linear relationship between resistance and temperature within the -10 °C to 60 °C range, with a temperature coefficient of resistance of 0.22 %. When the voltage increases, the temperature rise due to the EC effect, is directly measured by the thermistor. The temperature is extrapolated through a thermal model taking into account the sample shape, size, and thermal properties. Figure S 2 illustrates the temperature variation estimation in the case of a unipolar square voltage profile.

From the adiabatic temperature change and the $W_{loss}$ measured under unipolar square profile excitation, the cooling relative efficiency is computed using the following expression:

$$\eta_{cooling} = \frac{1}{1 + \frac{4TW_{loss}}{c\Delta T_{adia}^2}} \quad (2)$$

where $T$ is the working temperature and $c$ the sample specific heat capacity [11].

Based on temperature and calorimetry measurements of P(VDF-TrFE-CFE), it has been demonstrated that the heat capacity shows minimal dependence on the electric field. Therefore, consistent results can be achieved by assuming a constant heat capacity [31]. Using these findings as a foundation, the zero-field heat capacity has been employed.

The DSC (Differential scanning calorimetry) thermograms have been obtained using a T.A instruments Q20 apparatus, equipped with an intercooler, and subjected to N$_2$ flux. The experiments were carried over a temperature range spanning from -50 °C to 200 °C, employing a heating rate of 20 °C per minute. X-ray analysis were carried on thin terpolymer films (24 µm) at room temperature in transmission mode using a Brucker apparatus (Brucker, 2D phaser) with CuKα radiation of 1.79 A°. The sample-to-detector distance has been set at 10 cm. Diffraction spectra were measured at an angle range (2θ) between 5 and 30°.

## V. Results and discussion

Figure 3 illustrates the dielectric energy losses ($W_{loss}$) as a function of the peak field measured under sinusoidal unipolar excitation for samples deposited in N$_2$ and afterward submitted to different poling treatment (as cast, E poled, and E-T poled) compared with a non-treated reference sample prepared in air. As the electric field intensity increases, the energy losses in the film rise, in accordance with values reported in the literature for similar materials at the same frequency [20] ( Table S1 within the Supplemental Material [32]). This figure provides a basis for discussing three key results of the present work: firstly, the beneficial impact of both E, and E-T poling; ; secondly, the notable loss reduction observed on samples deposited under $N_2$ atmosphere.

1) The non poled film (blue curve) shows higher $W_{loss}$ compared to the poled films (black and red curves). At 50 V/µm, the loss of the non-poled film is approximately twice as high as that of the poled films, however this ratio decreases at higher field. The as-cast $N_2$ sample loss as a function of the peak field show a slight departure from linearity at fields above 40 V/µm. All the other samples losses, including the reference one (prepared in air and non-treated) show a remarkable non-linear behavior. Extrapolation up to 100 V/µm shows a rather similar behavior. This indicates that poling the film at an electric field higher than the one used for operation contributes to reducing losses in working conditions.

2) The electrically poled (red curve) and the electro-thermally poled (black curve) films show losses reduction. Up to 50 V/µm, electro-thermal poling shows the same benefit as E poling. On the contrary, when the excitation field gets above the field used for E-T poling (i.e. 50 V/µm) E poling, performed up to higher fields, is definitely more efficient in reducing losses. E poling to higher fields (130 V/µm) does not significantly change the loss reduction (Inset Figure 3), showing the existence of an upper bound for loss reduction through poling. Note that all the films studied here have large surfaces, a feature that is expected to increase the probability to get defect-induced breakdown. This limits the maximum electric field to be applied to the film to less than 130 V/µm.

The loss values measured in the poled films are lower than the ones reported in the literature on terpolymers to date, Table S1 within the Supplemental Material [32] (ref). For instance, the poled film shows losses that are 30% less than the ones reported by Qian [20] at 100V/µm.

3) The film deposited in $N_2$ atmosphere (red curve) shows lesser losses than the one deposited in air (green curve). Both have been E poled. Indeed, deposition in air due to a humidity of the order of 50%, led to a rougher film compared to the deposition in $N_2$ (Figure S 3. a). Roughness originates from water vapor-induced phase separation, a process that has been well-documented for the fabrication of PVDF membranes [33]. In the absence of water, no phase separation occurs, and films get smoother and denser. Due to humidity, phase separation or liquid–liquid demixing occurs resulting in rougher surfaces [33]. The high surface roughness of the film deposed in air entails field concentrations at the electrode/ terpolymer interface fostering charge injection, one of the main loss mechanisms .

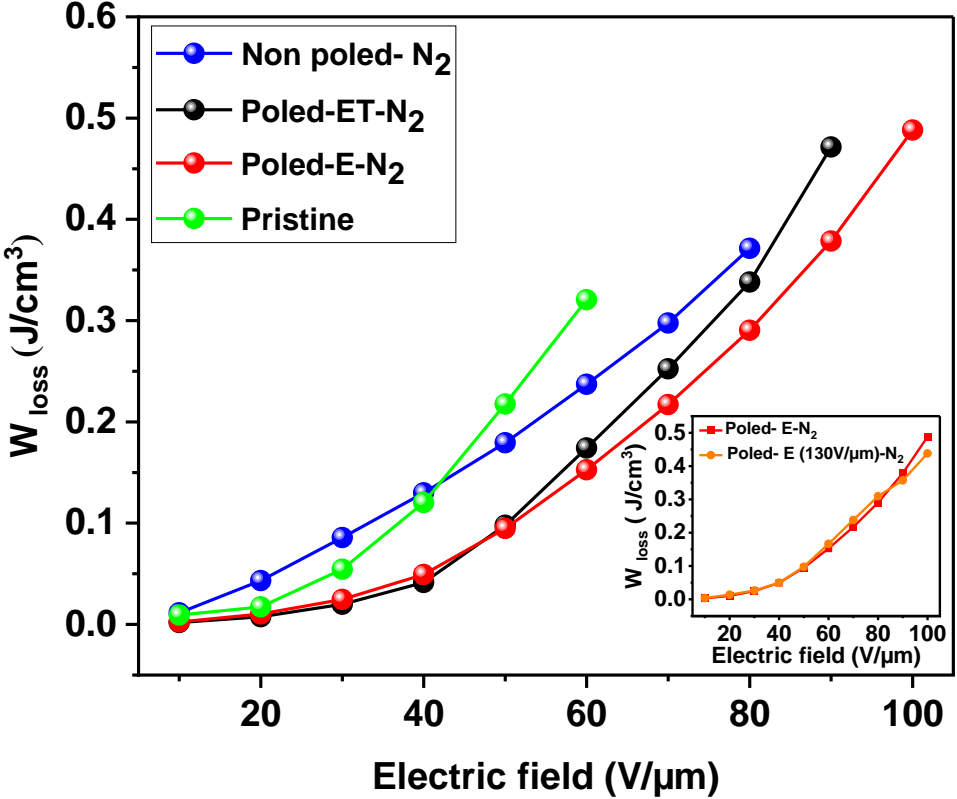

**Figure 3.** Loss as a function of the excitation peak electric field for samples fabricated under nitrogen ($N_2$) , as prepared (no-poling) and submitted to electrical (E) and electrothermal poling (ET). The green points represent the measurement carried on the reference sample (prepared in air with no further treatments). All samples are submitted to a unipolar sinusoid excitation at a frequency of 1 Hz. The inset shows a comparison of the losses for a film poled up to 100 V/µm and up to 130 V/µm.

To confirm the effect of the deposition conditions on the surface roughness, the surface profiles of the films have been measured. Figure S 3 compares the roughness measured on films fabricated under ambient air or under $N_2$ atmosphere. On substrate, film manufactured under conditions close to 0% humidity ($N_2$) exhibits a root-mean-square (rms) roughness of (~20 nm) and with an air humidity of the order of 50%, the rms roughness gets up to (~200 nm). Furthermore, we show that when the film is removed from the substrate, using a water droplet, new roughness at larger scale appears as apparent in Figure S 3.b. This additional roughness behaves as an oscillation perpendicular to the direction of peeling and is associated to the instability resulting from the adhesion force and the film elasticity during the substrate removal [34].

The charge/discharge energy, another feature relevant for applications, is shown in Figure 4. The main finding here is the 80% storage efficiency reached by the film deposited in $N_2$ atmosphere after poling, at 80-100 V/µm with an energy density of 1.30 J/cm$^3$ at 80 V/µm and 1.62 J/cm$^3$ at 100 V/µm. This is another significant improvement achieved through poling. This gets beyond similar results recently reported [35], where under same working conditions (i.e. 1.38 J/cm$^3$@100V/µm) a 65% storage efficiency has been achieved.

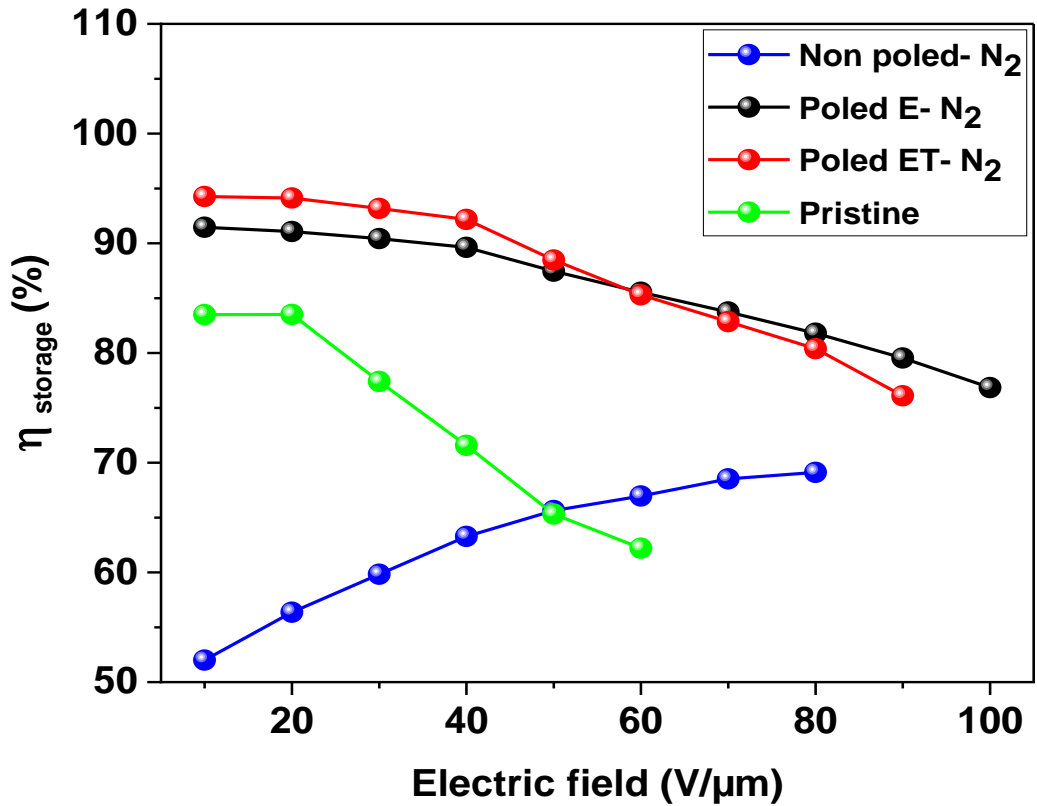

**Figure 4**. Efficiency of energy storage for deposition under nitrogen ($N_2$) and air, for non-poled, poled E, and poled E-T samples for sinus unipolar profile at a frequency of 1 Hz.

The adiabatic temperature change of poled and pristine films has been investigated as a function of the electric field using a direct temperature measurement. The $ΔT_{adia}$ values are very similar (~ 0.9°C) at moderate fields (below 40 V/µm). However, above 40 V/µm, they split, and the poled film shows a greater EC response than the other samples. $ΔT_{adia}$ For instance, at 60 V/µm, $ΔT_{adia}$ measured on the film poled in $N_2$ is 30% and 61% greater than the one measured on the non-poled and the poled in air respectively. There is 30% increase at 60 V/µm after film poling, and 61% compared to a film manufactured in the air (humidity). Above 90V/µm, the $ΔT_{adia}$ of the nitrogen prepared poled and non-poled films are quite similar.

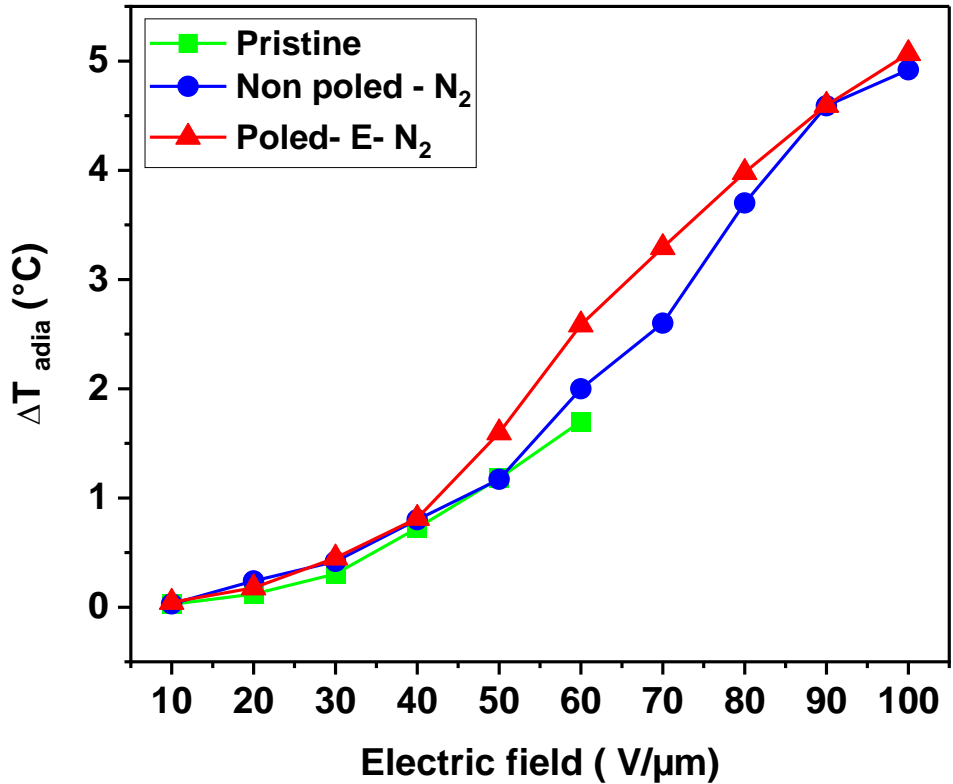

**Figure 5.** The evolution of adiabatic temperature as function of electric field for E- poled sample and non-poled one.

Despite being the two crucial features to assess the efficiency over a thermodynamic cycle, EC response and losses are rarely addressed at the same time (Table S1 within the Supplemental Material [32] (ref)).

A relevant figure of merit for cooling is the relative efficiency ($\eta_{cooling}$). It is calculated using Eq. (2), which can be deduced using the losses and the adiabatic temperature change.

The poled films (ET poled or E poled) show higher efficiency compared to the non-poled ones, confirming the results obtained from the energy storage characterization. Films with higher roughness exhibit lower efficiency confirming the smoothness of the surface as a key parameter for reaching higher efficiency.

The film showing the highest relative efficiency, up to 8% at 90 V/um, is electrically poled and deposited in $N_2$. The non-linear dependence of the efficiency as a function of the excitation field is due to the competition between adiabatic temperature change and losses. This results in a maximum of the efficiency as a function of the electric field. Namely, the film properties define an optimal working point for a given material.

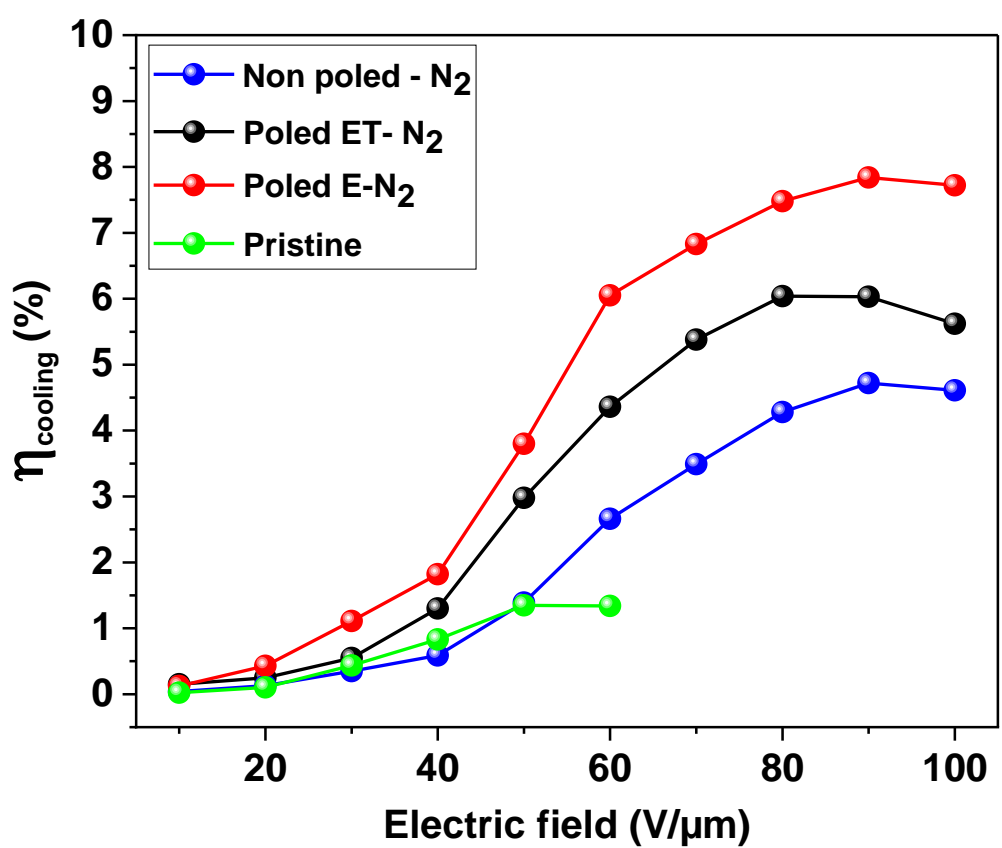

**Figure 6.** Maximum relative efficiency computed from Eq. (2) for the different films.

To investigate the effects of poling on the films, structural characterizations were conducted at various stages of the process, with a focus on XRD (X-ray diffraction) analysis at room temperature. Figure 7 shows the XRD data for the terpolymers near the [110, 200] reflection. Both non-poled and poled terpolymers exhibit a single intense peak around ~ 21°, indicating that the crystalline regions are in the nonpolar relaxor phase. The poled terpolymer shows a shift of the peak towards lower angles, accompanied by a narrowing and intensification of the peak.Interestingly. Assuming a two phases concept, the crystalline size (*L*) has been determined using the Scherrer formula.

$$L = 0.9\lambda/(\beta \cos\theta) \tag{3}$$

Here, $\lambda$ represents the X-ray wavelength, $\beta$ is the full width at half maximum (FWHM) of the peak, and $\theta$ is the Bragg angle. Using these parameters, the average crystalline sizes are calculated to be approximately 26 *nm* for non-poled and 44 *nm* for poled terpolymer Additionally, the crystallinity rate decreases with poling; the non-poled terpolymer has 37% crystallinity, while the poled terpolymer has only 33%.

The shift suggests a less correlated crystallite structure and an increased interplanar distance, measured at 4.89 Å for the poled sample compared to 4.86 Å for the pristine (non-poled) terpolymer.

After a duration of 2 weeks, no observable changes have been observed, indicating the stability of this new structure over time. In contrast, DSC measurements (Figure S 5) does not exhibit any discernible change (38.6% and 38.3%) in pristine, and poled sample.

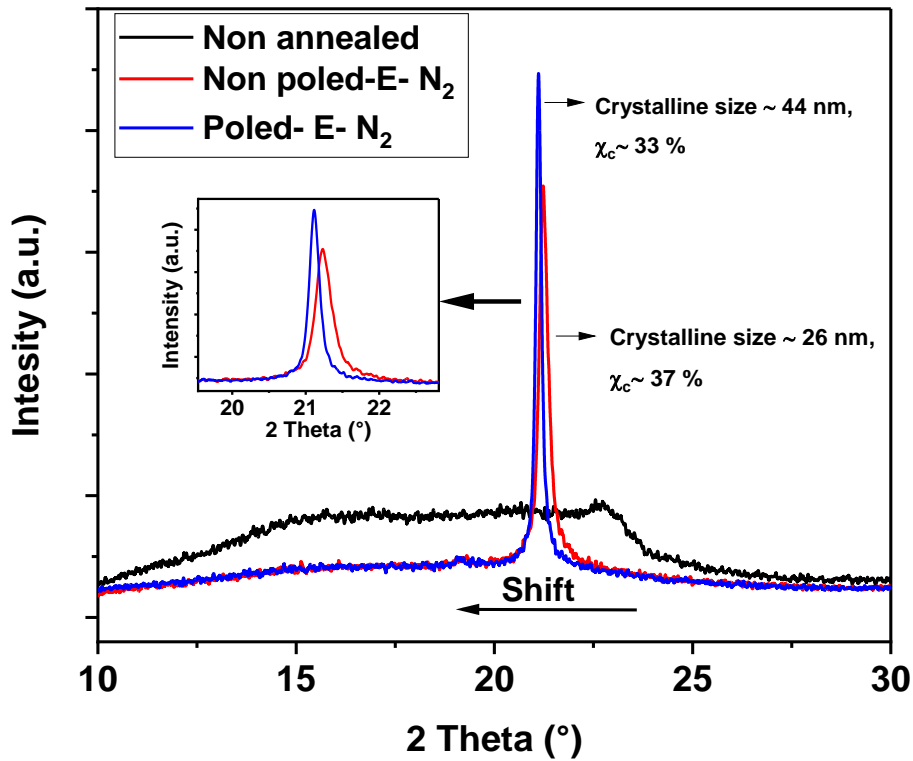

**Figure 7.** XRD measured on a not annealed sample, on an annealed not poled sample (black line) and on a poled sample (red line).

The effect of poling can be explained with the non-ergodicity of polymers, namely their non-unique configuration and the possibility to its tailoring driving the system to a state where its performances are improved.

After poling, although the polarization does not change appreciably, a larger EC response (i.e larger $\Delta T_{adia}$ /an increased entropy change) is observed.

Poling treatment may help relaxing residual stress due to the fabrication, and annealing processes, at the interface between the crystal, and the amorphous phases. However, this explanation remains largely conjectural, and further structural studies are necessary to validate it. The observed changes in the XRD peak profile could be attributed to crystal defects, such as micro-strain or strain gradients, rather than simply considering a two-phase concept (amorphous and crystalline without defects). This might also account for the different changes in crystallinity observed with DSC and XRD. It is worth noting that low-dissipation states induced by cyclic driving are a common feature in disordered matter, with similar behaviors reported in systems with diverse underlying microstructures, such as granular matter, polymers, and charge density waves in conductors [36].

A further insight can be obtained by comparing the film behavior P-E (Polarization-Electric Field) under the application of a positive (normal configuration) or a negative (inverted configuration) low electric field (+/-30 V/µm) (Figure S 4). In both cases losses are the same indicating a quite stable state. In addition, the negative field exhibits a slightly higher polarization, indicating a transition easier along the poled direction.

It is worth noting that here no asymmetry of losses with respect to the field direction has been observed. This rule out, for the case presented here, the electrode-interface ionic charges locking observed by Pedroli et al. [28] after a DC field electroannealing (i.e. after a rather different treatment than the one presented here). The terpolymer exhibits Relaxor Ferroelectric (RFE), Defect Ferroelectric (DFE), and Ferroelectric (FE) phases from low to large angles respectively, as referenced in studies [18] [17] [20] ,. It has been shown that an electric field induces a reversible transformation from the RFE phase to the DFE phase, and then to the FE phase that is from low to high angle. And, interestingly, XRD measurements at zero field after unipolar cycling revealed an expansion of the interplane distance (lower angle).

This behavior contrasts sharply with the increased maximum polarization and losses reported by Young et al. [21] following bipolar field excitation. This makes the behavior reported in this work rather different from the increasing of the maximum polarization, and of the losses reported by Young et al. [21] after cycling under bipolar field excitation.

Therefore, in spite of some similarity, neither the results reported in [28], nor the ones presented in [21] are expected to be associated with the same underlying mechanism driving the properties changes described in this work.

## VI. Electrocaloric adiabatic temperature change and efficiency of the optimized film

Finally, using the characterization protocol we recently presented [11], the adiabatic temperature change, and the dielectric losses are measured under relevant working conditions (voltage profile, frequency, peak electric field), over films showing the geometry (size, thickness) to be used in a typical EC cooling device. Both features, $\Delta T_{adia}$ and cooling efficiency extrapolated from the losses, are key properties for applications, as well as for the fundamental understanding of the order/disorder transition whence the EC effect originates.

Figure 8 shows the adiabatic temperature change of poled E-N$_2$ films in response to an applied electric field directly measured with the aforementioned thermistor setup. Measurements show a nonlinear relationship between EC response and the electric field. At low electric fields, the slope is small, then at 40 V/$\mu$m curves become steeper. The obtained $\Delta T_{adia}$ values stand out among the best performing EC outcomes recorded through direct measurements at moderate electric fields and on a non-modified terpolymer. (Table S1 within the Supplemental Material [32] compares the values presented here with those from the literature.)

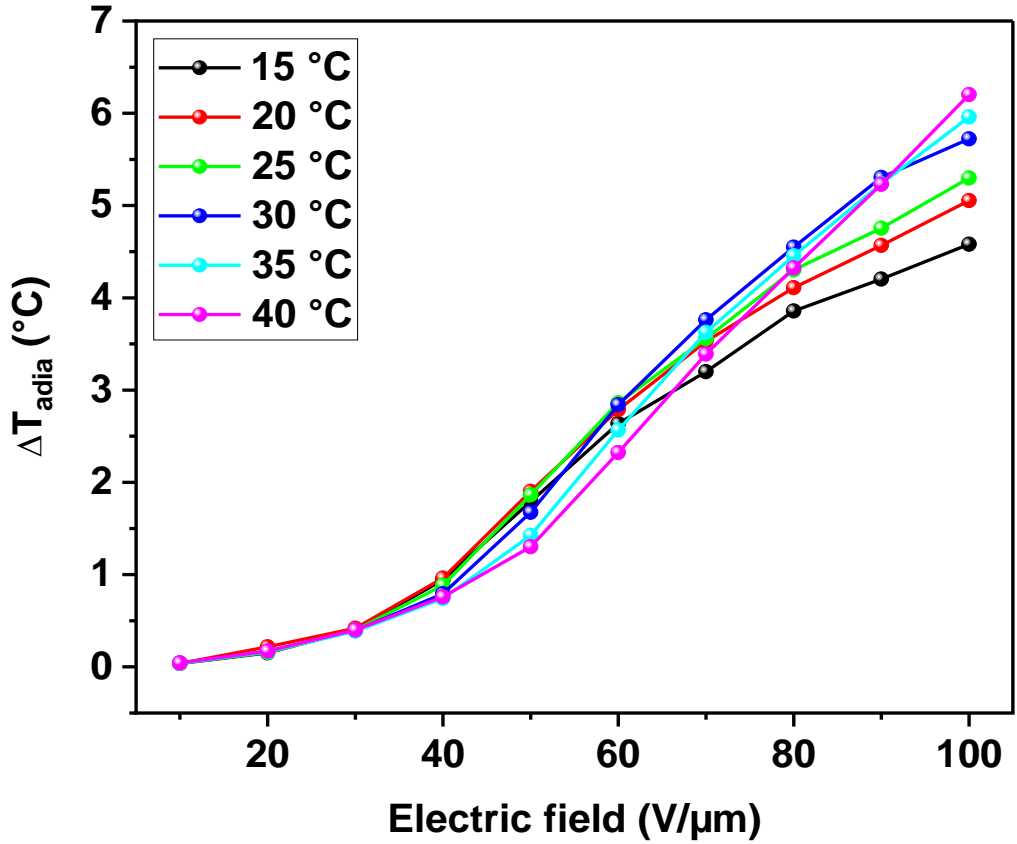

**Figure 8.** The evolution of the $\varDelta T_{adia}$ versus electric field at different temperatures for poled E-N$_2$.

Figure 9 shows the adiabatic temperature change at different working temperatures and for different electric fields.

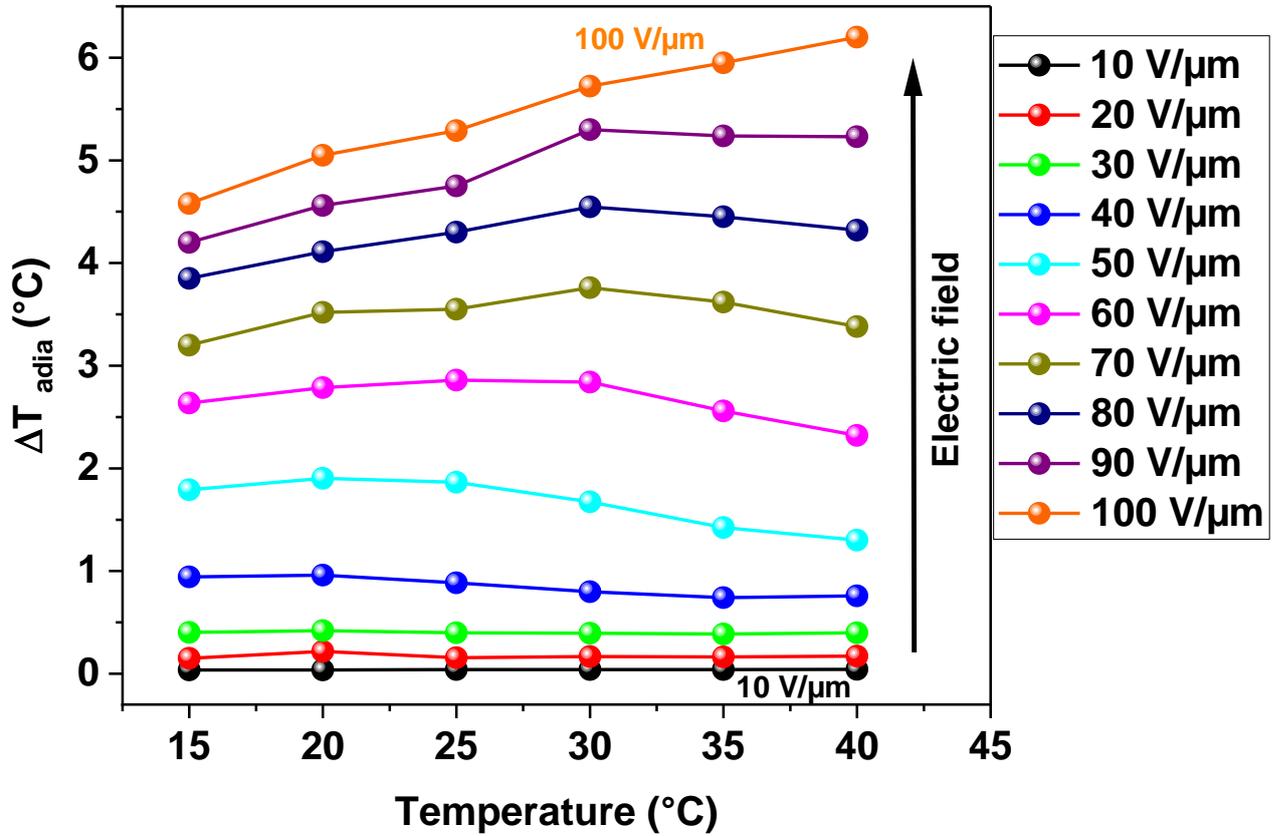

**Figure 9**. $\Delta T_{adia}$ versus different stage temperatures at different electric field of poled E-N$_2$

The relative efficiency ($\eta_{cooling}$) shown in Figure 10, offers an insight into the potential system efficiency across different operating points as influenced by the working temperature and the applied field. The curve illustrates that the ($\eta_{cooling}$) increases with temperature, until it reaches a maximum, beyond which it decreases as the temperature becomes higher. The temperature at which the efficiency reaches its maximum corresponds to the ambient temperature in the case of the terpolymer, with +-10°C variation, $\eta_{cooling}$ lost only 2-3%. If we go beyond 80 V/um, $\eta_{cooling}$ decreases. This is due to the significant losses increasing at high field.

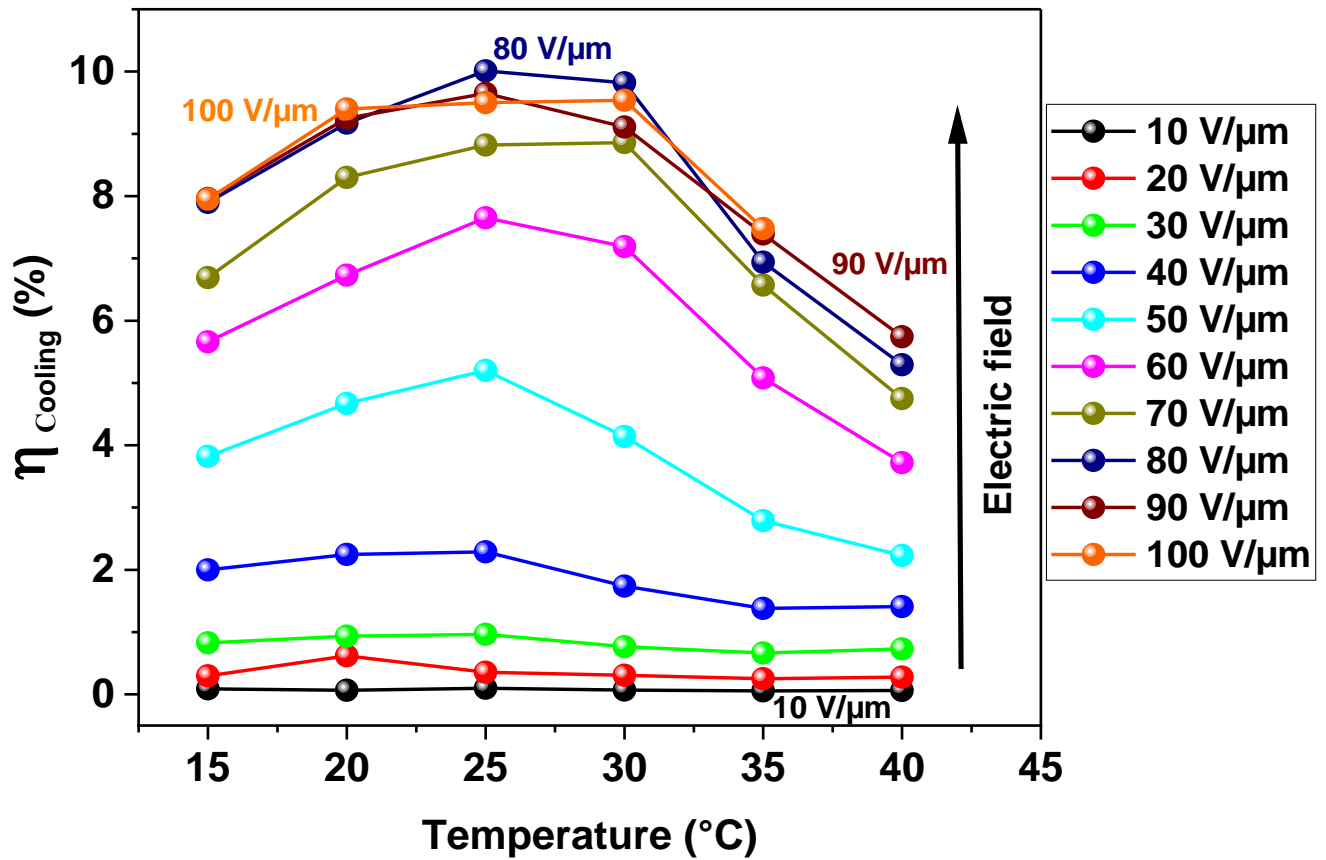

**Figure 10.** Efficiency ($\eta_{cooling}$) evolution versus temperature at difference electric fields (poled E-N$_2$)

Material deploying into a refrigeration device requires a high level of endurance. A single experiment comparing a "non-poled" and a "poled" film with a field endurance up to 60 V/μm under 1 Hz square unipolar excitation (i.e. typical device working conditions) shows a relevant improvement for the poled film. The poled film withstands 5 hours (17,700 cycles) of consecutive cycling, compared to the not-poled film which lasts for only 55 minutes. Besides, during our experiment we observed that non-poled films tend to break sooner than the poled ones.

**Conclusions**

This work presents two new methods to improve the properties of EC films for cooling applications. The first method achieves a notable loss reduction through film deposition under controlled atmosphere ($N_2$). This improvement is due to the reduction of local field concentration at the terpolymers/electrode interface associated with a smoother film surface.

The second method achieves a moderate improvement of the EC response, and a significant reduction of the dielectric losses through film poling under a progressively increasing unipolar electric field. The loss reduction is associated with the structural change fostered by poling (i.e. the expansion of the inter-plan distance observed by XRD). The expected cooling efficiency of the poled film, extrapolated from the direct measurement of the adiabatic temperature change, and of the dielectric losses, is expected to be improved from around 1% of the Carnot efficiency in the pristine samples up to 10% .

Furthermore, in relevant operating condition, the EC characterization of the optimized terpolymer has been presented for different electric field and at different temperatures. The films show an efficiency (η $_{cooling}$) up to 10% at 80 V/$\mu m$. Working at 100 V/$\mu m$, put $\Delta T_{adia}$ up to 5 °C but lower slightly the efficiency.

The results and the methods presented in this work pinpoint the complex relationship between $\Delta T_{adia}$ and dielectric losses in EC materials as the main challenge to reach the efficiency needed to implement the new generation of EC cooling devices making EC solid state refrigerators a mature technology.

See the Supplemental Material at (ref) for detailed data on direct EC measurements at 60 V/µm from various studies, polarization-electric field behavior under unipolar sinusoidal voltage, adiabatic temperature variation at 50 V/µm, and film profiles deposited in different atmospheres. Additionally, the material includes the evolution of energy losses over 10 cycles under square voltage and differential scanning calorimetry (DSC) thermograms showing thermal transitions ($T_{O-D}$ and $T_m$).

All data of the poled film deposited in $N_2$ are available in open access on opidor platform, and others are available on demand.

**References**

[1]  X. Ren, N. Meng, L. Ventura, S. Goutianos, E. Barbieri, H. Zhang, H. Yan, M. J. Reece, and E. Bilotti, *Ultra-High Energy Density Integrated Polymer Dielectric Capacitors*, J. Mater. Chem. A 10171 (2022).

[2]  R. Guo, H. Luo, D. Zhai, Z. Xiao, H. Xie, Y. Liu, X. Zhou, and D. Zhang, *Bilayer Structured PVDF-Based Composites via Integrating BaTiO3 Nanowire Arrays and BN Nanosheets for High Energy Density Capacitors*, Chem. Eng. J. **437**, 135497 (2022).

[3]  Y. Wang, J. Zhang, H. Gao, Y. Yao, C. Pu, J. Wang, L. Ren, and Q. Zong, *The Interface Effects on the Breakdown Strength of Multilayer PVDF-Based Capacitors*, Polymer (Guildf). **270**, 125803 (2023).

[4]  J. H. Bae and S. H. Chang, *PVDF-Based Ferroelectric Polymers and Dielectric Elastomers for Sensor and Actuator Applications: A Review*, Funct. Compos. Struct. **1**, 012003 (2019).

[5]  X. Chen, X. Han, and Q. D. Shen, *PVDF-Based Ferroelectric Polymers in Modern Flexible Electronics*, Adv. Electron. Mater. **3**, 1600460 (2017).

[6]  Q. Li et al., *Low-k Nano-Dielectrics Facilitate Electric-Field Induced Phase Transition in High-k Ferroelectric Polymers for Sustainable Electrocaloric Refrigeration*, Nat. Commun. **15**, 702 (2024).

[7]  Y. Meng, Z. Zhang, H. Wu, R. Wu, J. Wu, H. Wang, and Q. Pei, *A Cascade*

*Electrocaloric Cooling Device for Large Temperature Lift*, Nat. Energy **5**, 996 (2020).

[8] Y. Bo et al., *Electrostatic Actuating Double-Unit Electrocaloric Cooling Device with High Efficiency*, Adv. Energy Mater. **11**, 1 (2021).

[9] R. Ma, Z. Zhang, K. Tong, D. Huber, R. Kornbluh, Y. S. Ju, and Q. Pei, *Highly Efficient Electrocaloric Cooling with Electrostatic Actuation*, Science (80-. ). **357**, 1130 (2017).

[10] B. Neese, B. Chu, S. G. Lu, Y. Wang, E. Furman, and Q. M. Zhang, *Large Electrocaloric Effect in Ferroelectric Polymers near Room Temperature*, Science (80-. ). **321**, 821 (2008).

[11] N. Zeggai, B. Dkhil, M. Lobue, and M. Almanza, *Cooling Efficiency and Losses in Electrocaloric Materials*, Appl. Phys. Lett. **122**, 081903 (2023).

[12] Q. Zhang, Relaxor Terpolymers for Energy Storage Capacitors, 2005.

[13] P. Bai, Q. Zhang, H. Cui, Y. Bo, D. Zhang, W. He, Y. Chen, and R. Ma, *An Active Pixel-Matrix Electrocaloric Device for Targeted and Differential Thermal Management*, Adv. Mater. **2209181**, 1 (2023).

[14] X. Moya, E. Defay, V. Heine, and N. D. Mathur, *Too Cool to Work*, Nat. Phys. **11**, 202 (2015).

[15] J. Schipper, D. Bach, S. Mönch, C. Molin, and S. Gebhardt, *On the Efficiency of Caloric Materials in Direct Comparison with Exergetic Grades of Compressors*, J. Phys. Energy **5**, 045002 (2023).

[16] Y. Liu, B. Zhang, W. Xu, A. Haibibu, Z. Han, W. Lu, J. Bernholc, and Q. Wang, *Chirality-Induced Relaxor Properties in Ferroelectric Polymers*, Nat. Mater. **19**, 1169 (2020).

[17] F. Le Goupil, F. Coin, N. Pouriamanesh, and G. Fleury, *Electrocaloric Enhancement Induced by Cocrystallization of Vinylidene Difluoride-Based Polymer Blends .*, **10**, 1555 (2021).

[18] N. Zeggai, M. Fricaudet, N. Guiblin, B. Dkhil, M. Lobue, and M. Almanza, *UV-A-Irradiated P(VDF-TrFE-CFE) as a High-Efficiency Refrigerant for Solid-State Cooling*, ACS Appl. Polym. Mater. **6**, 3637 (2024).

[19] Y. Hambal, Electrocaloric Effect & Electrical Energy Storage in Ferroelectric Relaxor Polymers & Their Nanocomposites, Duisburg-Essen, 2022.

[20] X. Qian et al., *High-Entropy Polymer Produces a Giant Electrocaloric Effect at Low Fields*, Nature **600**, 664 (2021).

[21] L. Yang, E. Allahyarov, F. Guan, and L. Zhu, *Crystal Orientation and Temperature Effects on Double Hysteresis Loop Behavior in a Poly(Vinylidene Fl Uoride- Co -Tri Fl Uoroethylene- Co - Chlorotri Fl Uoroethylene)- Graf t -Polystyrene Graft Copolymer*, Macromolecules **46**, 9698 (2013).

[22] F. Le Goupil, K. Kallitsis, S. Tencé-Girault, C. Brochon, E. Cloutet, G. Fleury, and G. Hadziioannou, *Double-Bond-Induced Morphotropic Phase Boundary Leads to Enhanced Electrocaloric Effect in VDF-Based Polymer Flexible Devices*, ACS Appl. Energy Mater. **6**, 12172 (2023).

[23] S. Zheng et al., *Colossal Electrocaloric Effect in an Interface-Augmented Ferroelectric Polymer*, Science (80-. ). **382**, 1020 (2023).

[24] Fu-Chien Chiu, *A Review on Conduction Mechanisms in Dielectric Films*, Adv. Mater. Sci. Eng. **2014**, (2014).

[25] F. Pedroli, A. Marrani, M. Q. Le, O. Sanseau, P. J. Cottinet, and J. F. Capsal, *Reducing Leakage Current and Dielectric Losses of Electroactive Polymers through Electro-Annealing for High-Voltage Actuation*, RSC Adv. **9**, 12823 (2019).

[26] N. Pouriamanesh et al., *Limiting Relative Permittivity ' Burn-in ' in Polymer Ferroelectrics via Phase Stabilization*, ACS Macro Lett. **11**, 410 (2022).

[27] P. Liu, T. Zhang, C. Zhang, Y. Zhang, Y. Feng, Y. Zhang, Q. Chi, and C. Li, *Improved*

*Energy Storage Performance of P (VDF-TrFE-CFE) Films by Growing Superficial AlN Insulation Layer*, J. Mater. Sci. Mater. Electron. **34**, 1 (2023).

[28] F. Pedroli, A. Flocchini, A. Marrani, M. Q. Le, O. Sanseau, P. J. Cottinet, and J. F. Capsal, *Boosted Energy-Storage Efficiency by Controlling Conduction Loss of Multilayered Polymeric Capacitors*, Mater. Des. **192**, 108712 (2020).

[29] M. Q. Hoang et al., *Modelling the Impact of Electrode Roughness on Net Charge Density in Polyethylene*, J. Phys. D. Appl. Phys. **54**, 305303 (2021).

[30] M. Li, I. Katsouras, C. Piliego, G. Glasser, I. Lieberwirth, P. W. M. Blom, and D. M. De Leeuw, *Controlling the Microstructure of Poly(Vinylidene-Fluoride) (PVDF) Thin Films for Microelectronics*, J. Mater. Chem. C **1**, 7695 (2013).

[31] G. Sebald, L. Seveyrat, J. F. Capsal, P. J. Cottinet, and D. Guyomar, *Differential Scanning Calorimeter and Infrared Imaging for Electrocaloric Characterization of Poly(Vinylidene Fluoride-Trifluoroethylene- Chlorofluoroethylene) Terpolymer*, Appl. Phys. Lett. **101**, 022907 (2012).

[32] See the Supplemental Material at (URL will be inserted by publisher) for Detailed data on direct EC measurements at 60 V/µm from various studies, polarization-electric field response under unipolar sinusoidal voltage profile, adiabatic temperature variation at 50 V/µm, film profiles deposited in different atmospheres, evolution of energy losses over 10 cycles in a square voltage profile, and DSC thermograms showing TO-D and Tm thermal transitions.

[33] M. Li, I. L. Ilias Katsouras, Claudia Piliego, Gunnar Glasser, and P. W. M. B. and D. M. de Leeuw, *Controlling the Microstructure of Poly(Vinylidenefluoride) (PVDF) Thin Films for Microelectronics*, J. Mater. Chem. C **1**, 7645 (2013).

[34] A. Ghatak and M. K. Chaudhury, *Adhesion-Induced Instability Patterns in Thin Confined Elastic Film*, Langmuir 2621 (2003).

[35] Y. Hambal, V. V. Shvartsman, D. Lewin, C. H. Huat, X. Chen, I. Michiels, Q. Zhang, and D. C. Lupascu, *Effect of Composition on Polarization Hysteresis and Energy Storage Ability of p(Vdf-Trfe-Cfe) Relaxor Terpolymers*, Polymers (Basel). **13**, 1 (2021).

[36] N. C. Keim, J. D. Paulsen, Z. Zeravcic, S. Sastry, and S. R. Nagel, *Memory Formation in Matter*, Rev. Mod. Phys. **91**, 35002 (2019).

**Acknowledgements**

This work has benefited from the financial support of the LabEx LaSIPS (ANR-10-LABX-0032-LaSIPS) under the "Investissements d'avenir" program (ANR-11-IDEX-0003) and from (ANR-20-CE05-0044), which are both managed by the French National Research Agency.

**Declarations**

**Conflict of Interests/Competing Interests**

The authors declare that they have no known competing financial interests or personal relationships that could have appeared to influence the work reported in this paper.